\documentclass[english]{article}
\usepackage[T1]{fontenc}
\usepackage[latin9]{inputenc}
\usepackage{geometry}
\usepackage{babel}
\usepackage{array}
\usepackage{float}
\usepackage{url}
\usepackage{multirow}
\usepackage{amstext}
\usepackage{graphicx}
\usepackage[authoryear]{natbib}
\usepackage[unicode=true,pdfusetitle,
 bookmarks=true,bookmarksnumbered=false,bookmarksopen=false,
 breaklinks=false,pdfborder={0 0 0},pdfborderstyle={},backref=false,colorlinks=false]
 {hyperref}

\makeatletter

\providecommand{\tabularnewline}{\\}

\newcommand{\lyxaddress}[1]{
	\par {\raggedright #1
	\vspace{1.4em}
	\noindent\par}
}

\makeatother

\begin{document}
\title{Listening to the magnetosphere: How best to make ULF waves audible}
\author{Martin O. Archer\,$^{1,*}$, Marek Cottingham\,$^{1}$, Michael
D. Hartinger\,$^{2}$, Xueling Shi\,$^{3,4}$, Shane Coyle\,$^{3}$,\\
Ethan ``Duke'' Hill\,$^{3}$, Michael F. J. Fox\,$^{1}$, Emmanuel
V. Masongsong\,$^{5}$}
\maketitle

\lyxaddress{$^{1}$ Department of Physics, Imperial College London, London, United
Kingdom\\
$^{2}$ Space Science Institute, Boulder, Colorado, USA\\
$^{3}$ Department of Electrical and Computer Engineering, Virginia
Polytechnic Institute and State University, Blacksburg, Virginia,
USA\\
$^{4}$ High Altitude Observatory, National Center for Atmospheric
Research, Boulder, Colorado, USA\\
$^{5}$ Earth, Planetary, and Space Sciences Department, University
of California, Los Angeles, California, USA}
\begin{abstract}
Observations across the heliosphere typically rely on in situ spacecraft
observations producing time-series data. While often this data is
analysed visually, it lends itself more naturally to our sense of
sound. The simplest method of converting oscillatory data into audible
sound is audification -- a one-to-one mapping of data samples to
audio samples -- which has the benefit that no information is lost,
thus is a true representation of the original data. However, audification
can make some magnetospheric ULF waves observations pass by too quickly
for someone to realistically be able to listen to effectively. For
this reason, we detail various existing audio time scale modification
techniques developed for music, applying these to ULF wave observations
by spacecraft and exploring how they affect the properties of the
resulting audio. Through a public dialogue we arrive at recommendations
for ULF wave researchers on rendering these waves audible and discuss
the scientific and educational possibilities of these new methods.
\end{abstract}

\section{Introduction}

Ultra-low frequency (ULF) waves, with periods between seconds and
tens of minutes, provide a mechanism for solar wind energy and momentum
to be transferred into/around a planetary magnetosphere and couple
the space environment to the body's ionosphere. They are routinely
recorded as time-series data both by ground-based instruments such
as magnetometers or radar, as well as through in situ spacecraft observations
of the magnetospheric environment. Dynamic pressure variations, either
embedded in the large-scale solar wind or associated with meso-scale
kinetic structures at the bow shock, may excite any of the three magnetohydrodynamic
(MHD) waves -- the surface, fast magnetosonic, and Alfv\'{e}n modes
-- at locations within the magnetosphere \citep[e.g.][]{sibeck89,hartinger13}.
In addition, MHD or other plasma waves in the ULF range -- such as
electromagnetic ion cyclotron (EMIC) and mirror modes -- may also
be generated internally through fluid/kinetic instabilities or wave-particle
interactions \citep[e.g.][]{cornwall65,southwood69,hasegawa75}. All
of these different waves may exist as incoherent broadband enhancements
in wave power or at well-defined discrete, or even an entire spectrum,
of coherent oscillations. Thus a zoo of ULF wave phenomena are known
to occur within Earth's magnetosphere (see the review of \citealp{nakariakov16},
for more).

Classification of magnetospheric ULF waves \citep{jacobs64} has long
been separated into whether pulsations qualitatively are quasi-sinusoidal
(continuous pulsations, Pc) or more irregular (Pi). These two categories
have then be subdivided into near-logarithmically spaced frequency
bands. Unfortunately, such a classification scheme does not take into
account the physical processes involved in the generation and propagation
of the waves, nor how these may be reflected in their physical or
morphological properties. Indeed, many studies simply consider the
integrated power over one or several of these frequency bands \citep[e.g.][]{mann04},
which will not distinguish between broadband, narrowband, or multi-harmonic
signals. It is known, however, that the frequencies of physically
different ULF waves may overlap and even occupy wildly different bands
depending on the conditions present \citep[e.g.][]{archer15b,archer17}.

Beyond classification, even the analysis of ULF wave measurements
can be quite challenging, since they are often highly nonstationary
and may exhibit nonlinearity. Time-frequency analysis which can capture
these variations are therefore required. Methods based on the linear
Fourier transform are most commonly used, though these spectral estimators
often result in a large amount of variance making it difficult to
distinguish true spectral peaks to simply a realisation of an underlying
stochastic process \citep{chatfield89}. While statistical tests have
been developed to help address this \citep[e.g.][]{dimatteo18,dimatteo21},
these are not immune to false positives (or indeed false negatives).
The continuous wavelet transform \citep{torrence98} offers some potential
improvements to Fourier methods in self-consistent time-frequency
analysis, e.g. enabling feature extraction. Wavelets are, like Fourier
methods, still subject to the Gabor uncertainty principle in time-frequency
space ($\sigma_{t}\sigma_{f}\geq\frac{1}{2}$, where $\sigma_{t}$
and $\sigma_{f}$ represent standard deviations in time and frequency
respectively) due to their linear nature. Similarly large variances
in spectral power again occur, which can limit the identification
of discrete frequency signals. Nonlinear time-frequency analysis methods,
which are not constrained to the Gabor limit, exist including the
Wigner-Ville distribution \citep{wigner32,ville48} and Empirical
Mode Decomposition \citep{huang98}. However, these are not currently
widely used for the analysis of ULF waves and their associated artifacts,
mode mixing, or stability in these applications are not fully understood
\citep{chi08,piersanti18}. Ultimately, few studies employ fully automated
detection, with visual inspection still often used either for identification
or confirmation of ULF wave signals in real data \citep[e.g.][]{takahashi15}.
However, oscillatory time-series data naturally lends itself most
to another of our senses -- sound.

Sonification refers to the use of non-speech audio to convey information
or perceptualise data \citep{kramer94}. The human auditory system
has many advantages over visualization in terms of temporal, spatial,
amplitude, and frequency resolution. For example the human hearing
range of 20--20,000~Hz spans three orders of magnitude in frequency
and at least 13 orders of magnitude in sound pressure level \citep{suzuki04},
compared to the human visual system's only a quarter order of magnitude
in wavelengths and no more than 4 orders of magnitude in luminance
\citep{kunkel10}. Human hearing is also highly nonlinear, hence is
not subject to the Gabor limit, thus can identify the pitch and timing
of sound signals much more precisely than linear methods \citep{oppenheim13}.
Furthermore, the human auditory system's ability to separate sounds
corresponding to different sources far outperforms even some of the
most sophisticated blind source separation algorithms developed \citep[e.g.][]{qian18}.

\begin{figure}[H]
\begin{centering}
\includegraphics{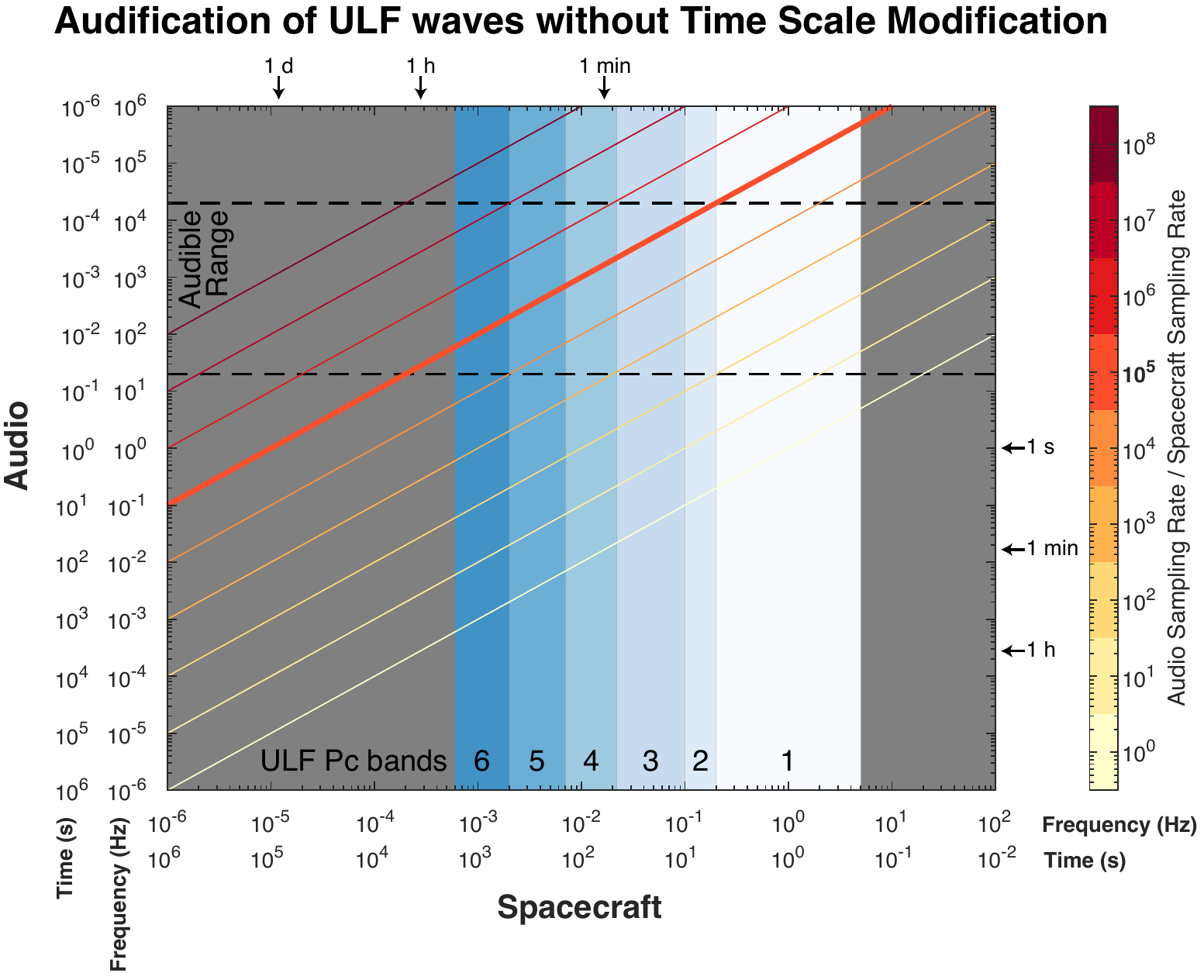}
\par\end{centering}
\caption{Relationship between spacecraft time and frequency to audio time and
frequency in audification for different ratios of sampling rates.
ULF wave bands are also highlighted.\label{fig:audification}}

\end{figure}

The simplest method of converting an oscillatory time-series into
sound is a one-to-one mapping of data samples to audio samples, known
as audification \citep[e.g.][]{alexander11,alexander14}. The benefit
of this technique is that no information is lost and the audio is
therefore a true representation of the original data. Other highly-used
sonification techniques require the mapping of data values (or the
outputs of some analysis on them) to discrete synthesised musical
notes \citep{bearman12,phillips19}. This necessarily loses some information
in the process and may also impose the user's desired aesthetics on
the data, meaning that it is arguable whether the audio is truly representative
of the underlying data. In direct audification the only free choices
are the amplitude to normalise the original data (audio is stored
as dimensionless values between -1 and +1) and the sampling rate with
which to output the audio. There is a straightforward relationship
between time durations in the audio and their corresponding durations
in the original data, given by

\begin{equation}
\Delta\left(\mathrm{Audio\,time}\right)=\Delta\left(\mathrm{Spacecraft\,time}\right)\times\frac{\mathrm{Spacecraft\,sampling\,rate}}{\mathrm{Audio\,sampling\,rate}}\label{eq:time-audification}
\end{equation}
, where the spacecraft time represents the actual time of the spacecraft
observations (e.g. in UTC) when events took place. Since frequency
is the reciprocal of the time period, audio and spacecraft frequencies
are related by

\begin{equation}
\mathrm{Audio\,frequency}=\mathrm{Spacecraft\,frequency}\times\frac{\mathrm{Audio\,sampling\,rate}}{\mathrm{Spacecraft\,sampling\,rate}}\label{eq:audio-freq}
\end{equation}
Figure~\ref{fig:audification} demonstrates these relationships,
indicating how various choices in the ratio of the audio to spacecraft
sampling rates renders different frequency ranges in the original
data audible. It is clear that for the Pc3--6 bands of ULF waves,
where MHD waves largely fall, then an audio to spacecraft sampling
rate ratio of order $10^{5}$ is required (corresponding to the thicker
line in the figure). Since space plasma missions typically produce
data with cadences of a few seconds, this means that typical audio
sampling rates (such as 44,100~Hz) may be used. The large ratio means
that timescales of the audio are dramatically reduced, which is advantageous
as data navigation, mining and analysis will have a reduced processing
time through listening \citep{hermann2002}.

\citet{alexander11,alexander14} and \citet{wicks16} showed that
researchers using audification applied to Wind magnetometer data in
the solar wind aided in the identification of subtle features present
that were not necessarily clear from standard visual analysis. \citet{archer18}
similarly showed that audification of GOES magnetic field observations
enabled high school students to identify numerous long-lasting decreasing-frequency
poloidal field line resonances during the recovery phase of geomagnetic
storms. Such events were previously thought to be rare, but through
exploration of the audio they were in fact demonstrated to be relatively
common. These preliminary studies into the use of audification of
the ULF wave bands within heliophysics show promise in its potential
scientific applications.

Direct audification may not, however, be suitable for all space plasma
missions and/or ULF wave events. Figure~\ref{fig:audification} indicates
that to make the ULF bands audible necessarily renders a day of data
into about one second of audio. In geostationary orbit the background
plasma and magnetic field conditions, which affect the eigenfrequencies
of the ULF modes, are relatively stable across the orbit \citep{takahashi10}.
This makes identifying the local time patterns in frequency and occurrence
of ULF waves relatively straightforward \citep{archer18}. However,
for more eccentric orbits with similar orbital periods the conditions,
frequencies, and occurrence of ULF waves will rapidly change throughout
the orbit \citep{archer15b,archer17}. Therefore, the rate at which
ULF waves' properties may be changing in the audio will be very fast.
This is demonstrated in the audio in Supplementary~Data~1 of \citet{suppdata},
where audification is applied to idealised and real events from the
THEMIS mission. It is clear from this audio that changes are occurring
too quickly to effectively listen to and analyse. Another potential
issue in audification is that ULF waves can be highly transient, occuring
for only several oscillations due to intermittent driving and/or damping
\citep[e.g.][]{zong09,wang15,archer19}. This means that ULF wave
events which persist for only $\sim10\text{--}30\,\mathrm{min}$ will
last only $\sim60\text{--}200\,\mathrm{ms}$ in the audio. Furthermore,
effective pitch discrimination by the auditory system often requires
several tens of (and even up to a hundred) oscillations \citep{fyk87},
which is not always the case with ULF waves. Thus, improvements to
the sonification process over simple audification are clearly required
for application to magnetospheric ULF waves more widely.

In this transdisciplinary paper we introduce several potential improved
sonification methods for ULF waves, borrowing techniques from the
fields of audio and music. The properties of the resulting audio from
these different methods are assessed through a public dialogue with
stakeholder groups to arrive at recommendations for ULF wave researchers
on how best to render these waves audible. Finally, we discuss the
scientific and educational possibilities that might be enabled by
these new methods.

\section{Sonification}

Taking existing techniques from the fields of audio and music, we
have developed new potential sonification methods for magnetospheric
ULF waves which we detail throughout this section. In this work we
focus on Alfv\'{e}n waves \citep{southwood74}, arguably the most
intensely studied mode of ULF waves (e.g. the review of \citealp{keiling16}).
Further justifying this choice is the fact that significant coupling
occurs within the magnetosphere from compressional to Alfv\'{e}n
modes due to plasma inhomogeneity, curvilinear magnetic geometry,
and hot plasma effects \citep{southwood84}. However, this focus does
not preclude that sonification might also render other types of ULF
waves that also occupy the same Pc3--6 frequency bands, such as surface
\citep[e.g.][]{archer19} or waveguide \citep[e.g.][]{hartinger13b}
modes, effectively audible, though these applications are beyond the
scope of this paper. The natural frequencies of Alfv\'{e}n waves
vary with local time and $L$-shell, with the spectrum of frequencies
with location being known as the Alfv\'{e}n continuum. The typically
reported trend is that, outside of the plasmasphere or plumes, Alfv\'{e}n
wave frequencies tend to decrease with radial distance from the Earth
\citep{takahashi15b}, hence spacecraft observations often show tones
whose frequencies sweep from high to low as the probe follows its
orbit from perigee to apogee (and vice versa as the orbit continues
from apogee back to perigee). It is worth noting though that the Alfv\'{e}n
continuum can vary considerably, both in terms of absolute values
and morphology, with solar wind and magnetospheric driving conditions
\citep{archer15b,archer17}.

\begin{figure}[tbph]
\begin{centering}
\includegraphics[width=1\columnwidth]{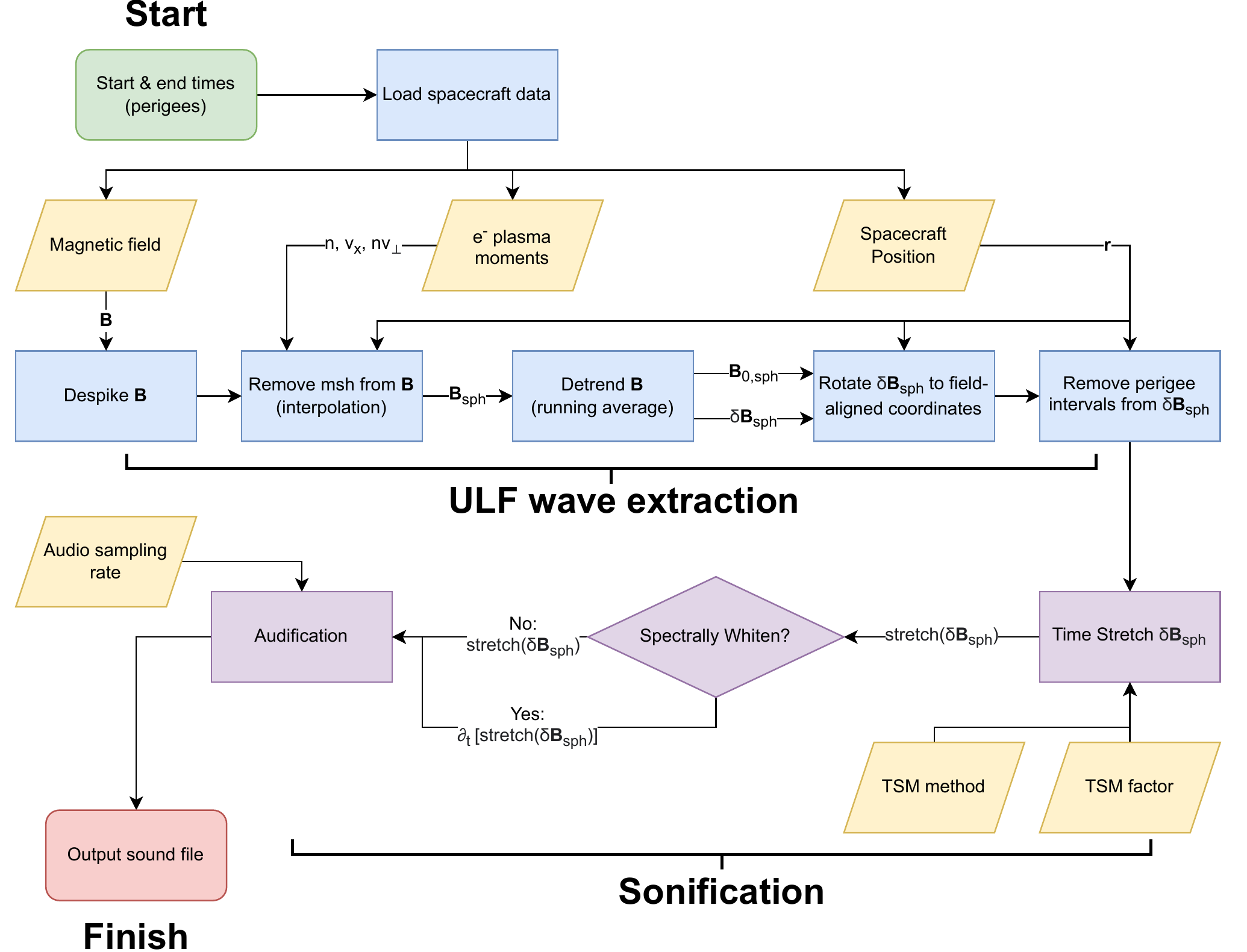}
\par\end{centering}
\caption{Flow chart of the sonification processes used.\label{fig:flow-chart}}

\end{figure}

The full sonification process developed here is outlined as a flow
chart in Figure~\ref{fig:flow-chart}. Throughout we apply the methods
to NASA THEMIS observations \citep{angelopoulos08}, chosen due to
the highly eccentric equatorial orbits of its spacecraft with the
inner three probes having apogees $r\sim12\text{--}15\,\mathrm{R_{E}}$
and perigee $r\sim1.5\,\mathrm{R_{E}}$ over the course of the mission.
Spin-resolution data is used, with one data point every $3\,\mathrm{s}$
(any data gaps or irregular samples are regularised by interpolation).
As with \citet{archer18} we focus only on waves in the magnetic field,
using fluxgate magnetometer data (\citealp{auster08}; electron plasma
data, \citealp{mcfadden08a}, is also used for discriminating between
magnetosphere and magnetosheath intervals). Other physical quantities
such as the plasma velocity \citep[e.g.][]{takahashi15b} or electric
field \citep[e.g.][]{hartinger12} could also be used in the sonification
of ULF waves -- a prospect we leave to future work. How the different
potential sonification methods affect the sound of the resulting audio
is assessed in section~\ref{sec:Public-dialogue}. The software developed
incorporating all of these methods is available at \url{https://github.com/Shirling-VT/HARP_sonification}.

\subsection{ULF wave extraction}

Before sonification, the magnetospheric ULF waves must be extracted
from the data and transformed into an appropriate coordinate system.
Non-physical spikes in the magnetometer data are first removed by
identifying when $\partial_{t}\mathbf{B}>3\,\mathrm{nT}\,\mathrm{s}^{-1}$
in any component and removing the 10 neighbour data points from all
components. Next, since magnetospheric ULF waves are the focus of
the sonification, magnetosheath intervals are also removed. These
are flagged for $r>8\,\mathrm{R_{E}}$ when either the electron number
density is greater than $10\,\mathrm{cm}^{-3}$, antisunward component
of the electron velocity is greater than $200\,\mathrm{km}\,\mathrm{s}^{-1}$,
or the perpendicular electron particle flux is greater than $2\times10^{7}\,\mathrm{cm}^{-2}\,\mathrm{s}^{-1}$.
The three nearest neighbour points to flagged data are also removed.
Interpolation is applied to fill in the removed data, thereby preventing
discontinuities in the data which might affect the sonification. For
short magnetosheath intervals, less than the dominant ULF wave period
present, the interpolation will fill in the gaps in phase effectively.
In contrast, when the intervals are longer than ULF wave periodicities
then the resulting audio during interpolated intervals will be silent,
since frequencies associated with the interpolation will be below
the audible range.

The background magnetospheric magnetic field is determined by a $30\,\mathrm{min}$
running average, with this trend being subtracted to arrive at the
ULF waves. These are then rotated into a field-aligned coordinate
system. The field-aligned component (representative of compressional
waves) is along the previous running average, the radial component
(representative of poloidal waves) is perpendicular to this pointing
radially away from Earth, and the azimuthal component (representative
of toroidal waves) is perpendicular to both of these pointing eastwards.
Since close to Earth the instrument precision is reduced \citep{auster08}
and the dipole field as measured by the spacecraft changes more rapidly
than the running average \citep{archer13b}, we finally remove all
data from geocentric distances $r<5\,\mathrm{R_{E}}$ setting these
points to zero.

\subsection{Time scale modification}

Time scale modification (TSM) refers to speeding up or slowing down
audio without affecting the frequency content, e.g. the pitch of any
tones \citep{driedger16}. It therefore modifies the link between
the time of events and frequency/periodicity of waves. This may be
advantageous for the sonification of ULF waves compared to simple
audification, since it allows the frequencies to still fall within
the audible range as per equation~\ref{eq:audio-freq} but stretches
the audio in time so that events do not occur so quickly for listening,
i.e.

\begin{equation}
\Delta\left(\mathrm{Audio\,time}\right)=\mathrm{TSM\,factor}\times\Delta\left(\mathrm{Spacecraft\,time}\right)\times\frac{\mathrm{Spacecraft\,sampling\,rate}}{\mathrm{Audio\,sampling\,rate}}\label{eq:time-tsm}
\end{equation}
Here the TSM factor refers to the factor by which the audio has been
stretched in time, where values greater than one result in longer
audio (in some other sources the definition may refer to the reciprocal
of this). Time stretching necessarily increases the number of oscillations
present in each event, since the periodicities are maintained and
thus equation~\ref{eq:audio-freq} remains unaffected. This has the
consequence of also improving the audibility of short-lived waves
for purposes such as pitch detection \citep{fyk87}.

One of the benefits of direct audification is that the resulting audio
is identical in information content to the original data. While TSM
methods necessarily modify their inputs, this is done in ways which
are relatively easy to understand. In principle, it should be possible
to reverse these procedures to arrive back at the original data, which
we justify for each method in turn. However, in practice some additional
artifacts may be present when attempting this reversal.Nonetheless,
key properties of the original data are left invariant by each process
and thus the time-stretched audio can still be treated as a representation
of the original data. Because different TSM methods work differently
though, it is expected that the methods will produce audio that sounds
different.

In this subsection we briefly describe several widely used TSM methods,
originally developed primarily for music, which we will apply to ULF
wave observations. For further discussion of the details behind these
methods see the review of \citet{driedger16} and/or our provided
software. Throughout this subsection the input refers to the extracted
ULF waves from the original spacecraft time series, consisting of
$N$ data samples (the spacecraft time range used multiplied by the
spacecraft sampling rate). Several of the methods are based on \textit{overlap--add}
procedures. These take \textit{analysis frames}, highly overlapping
windows spaced by the \textit{analysis hop}, from the input data.
Throughout we have set the length of the analysis frames to be 512
samples ($25\,\mathrm{min}$) as this is the closest power of two
to the length of the running average used to extract the waves. The
analysis frames are individually manipulated, depending on the TSM
method used, and then relocated on the audio time axis as \textit{synthesis
frames} with corresponding \textit{synthesis hop}. The slightly overlapping
synthesis frames are then added together, essentially performing multiple
copies of very similar parts of the original data, yielding the output.
This gives the desired TSM of the input data by a stretch factor equal
to the ratio of analysis to synthesis hops. The output thus contains
more data samples but covers the same spacecraft time range in UTC,
with the periodicities of oscillations (shorter than the frame length)
in terms of data samples having been preserved. In general one has
a freedom in what fraction of the frame length to use as the synthesis
hop. This has been tested and we report which choices seemed to yield
the best results when applied to magnetospheric ULF waves. Mathematically
it is possible to reproduce the original signal following an overlap-add
procedure with no modifications subject to some simple constraints
on the windowing function \citep{sharpe}. Therefore, if the modifications
are also invertible then overlap-add based TSM methods should also
be reversible. Figure~S1 shows an example of applying first stretching
and then compression of time for each TSM method to an idealised chirp
ULF wave signal with added noise.

\subsubsection{Waveform Similarity Overlap-Add (WSOLA)}

\begin{figure}[tbph]
\begin{centering}
\includegraphics{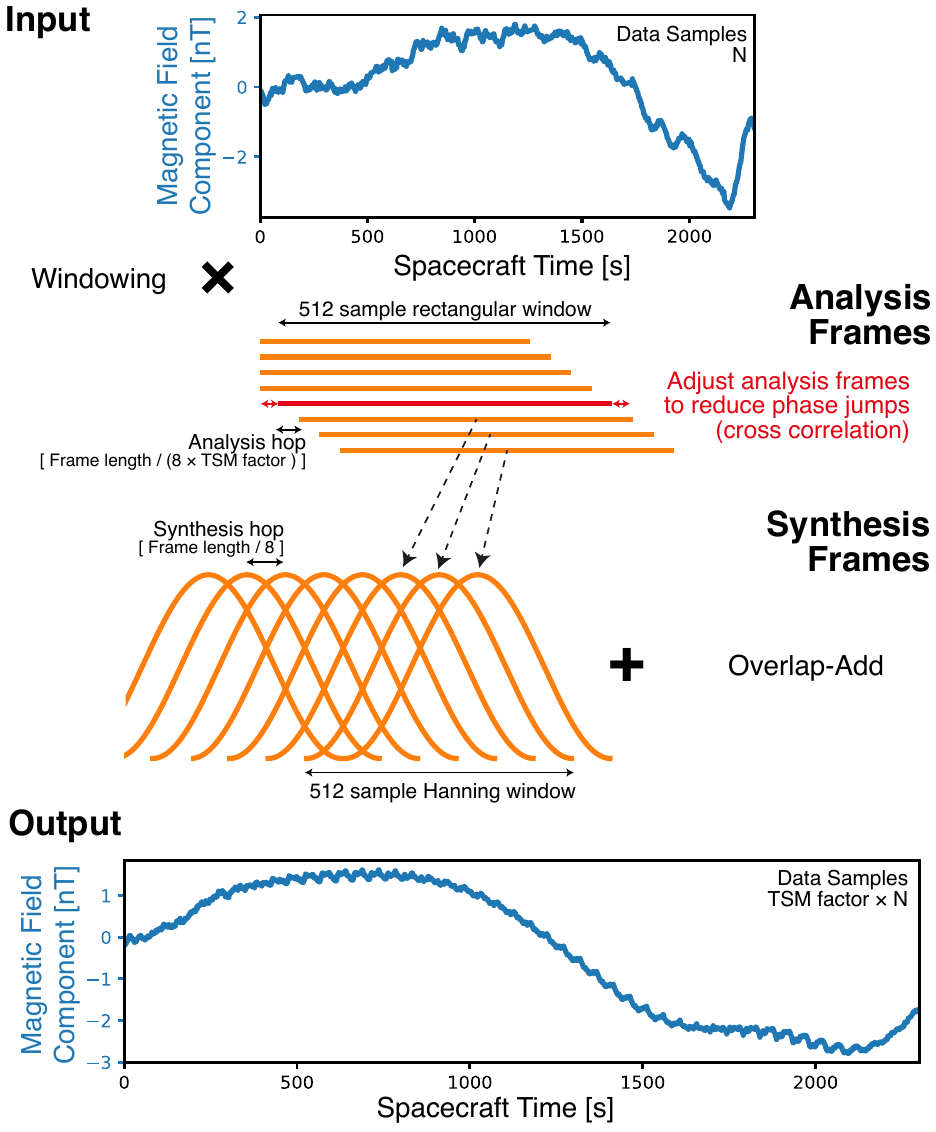}
\par\end{centering}
\caption{Illustration of the Waveform Similarity and Overlap Add (WSOLA) method.\label{fig:WSOLA}}

\end{figure}

Waveform Similarity Overlap-Add (WSOLA) is a TSM method that works
in the time domain as outlined in Figure~\ref{fig:WSOLA}. The only
modification between the analysis and synthesis frames are slight
shifts in time to reduce any phase jumps present in the output, with
these shifts being determined for each successive frame via cross
correlation \citep{driedger16}. As a time-based procedure WSOLA is
known to have issues with transients much shorter than the analysis
frame length, causing these features to be repeated in the output.
WSOLA is also known to struggle when multiple harmonic sources are
present, with only the largest amplitude one being preserved in the
output. For continuous functions with no noise, the cross-correlation
function should be smooth with a well-defined peak. Therefore, the
time-shifts applied in WSOLA are invertible in principle. However,
in practice the discrete-time nature of digital data and the presence
of substantial noise may hinder the invertibility of WSOLA. The WSOLA
example in Figure~S1 (blue) shows that the original periodicities
are recoverable, but amplitudes and longer-term trends may end up
being rather different.

\subsubsection{Phase vocoder}

\begin{figure}[tbph]
\begin{centering}
\includegraphics{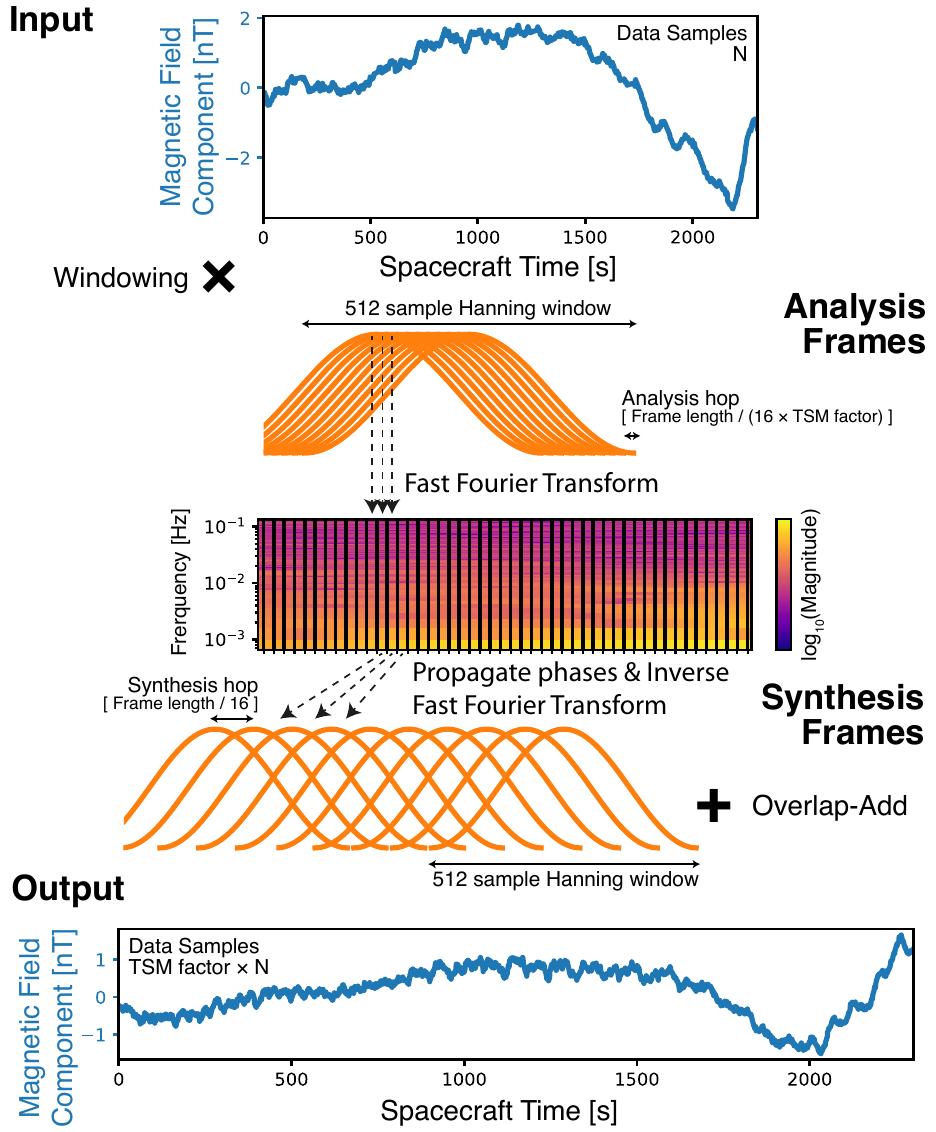}
\par\end{centering}
\caption{Illustration of the Phase Vocoder method.\label{fig:phase-vocoder}}

\end{figure}

The phase vocoder TSM is a frequency domain overlap--add procedure
that aims to preserve the periodicities of all signal components.
Figure~\ref{fig:phase-vocoder} depicts how it works through a Short
Time Fourier Transform (STFT). Due to the coarse nature of this transform
in both time and frequency, to maintain continuity in the output the
STFT phases at each frequency need to be iteratively adjusted (based
on the phase differences between frames) before inversion, known as
phase propagation \citep{driedger16}. Therefore, the STFT spectrogram
(absolute magnitude of the transform) remains invariant under the
phase vocoder (note human hearing is relatively insensitive to phase;
\citealp{meddia91,moore02}). Transient signals tend to get somewhat
smeared out over timescales of around the frame length using this
method. While locking the phase of frequencies near peaks in the spectrum
can reduce phase artifacts, these nonetheless may still be present
and are reported to have rather distinct sounding distortions. The
STFT is known to have an inverse transform, however, again in practice
the discrete application of this transform to noisy data can result
in artifacts such as time-aliasing \citep{allen77}. The example of
its invertibility in Figure~S1 (orange) demonstrates it performs
slightly better than WSOLA at recovering the original data.

\subsubsection{PaulStretch}

\begin{figure}[tbph]
\begin{centering}
\includegraphics{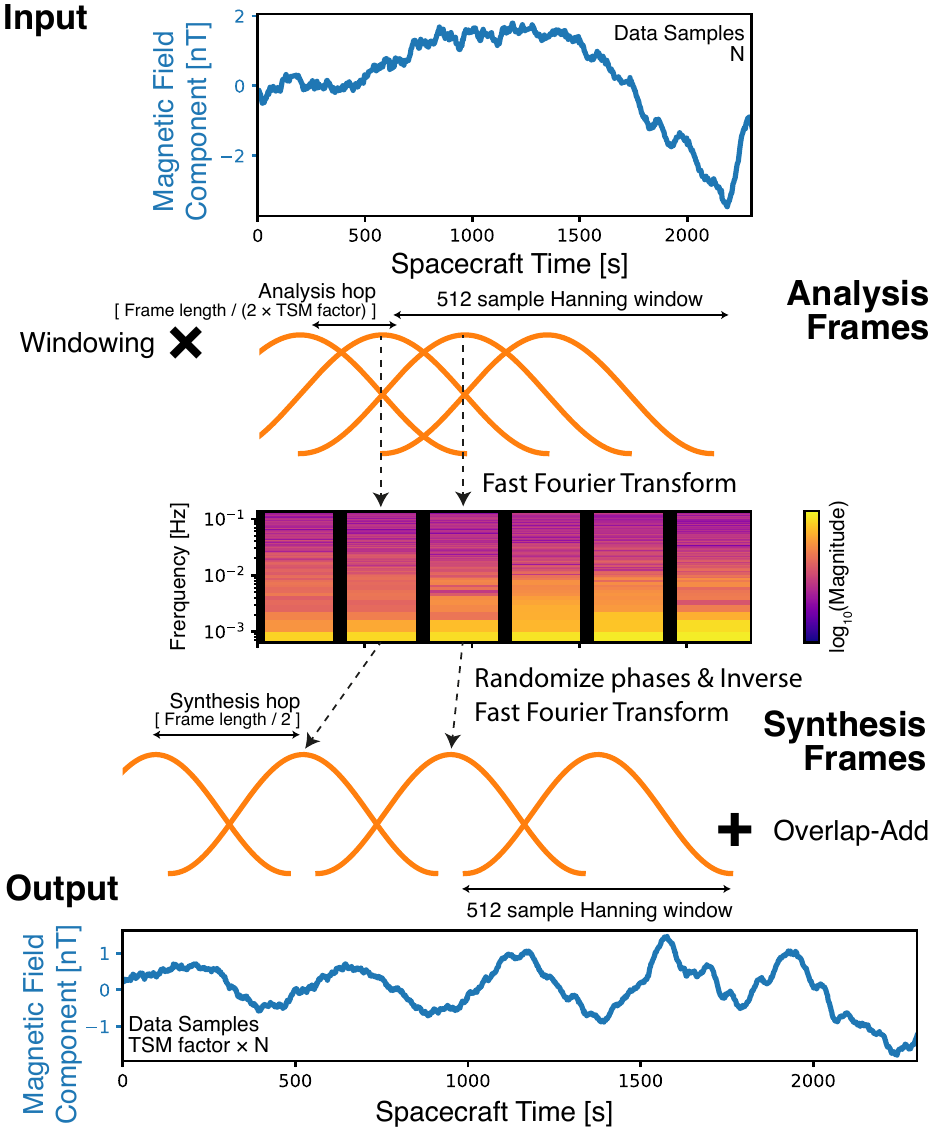}
\par\end{centering}
\caption{Illustration of the PaulStretch method.\label{fig:paulstretch}}

\end{figure}

PaulStretch is an extreme sound stretching algorithm based on the
phase vocoder \citep{paulstretch}. Instead of propagating the phase,
which becomes difficult for large TSM factors, the algorithm instead
randomises all the STFT phases, as shown in Figure~\ref{fig:paulstretch}.
This results in a more smooth sound than the phase vocoder method,
with less repetition and distortion \citep{paulstretch2}. Unlike
the other methods presented here, PaulStretch is not suitable for
TSM factors less than unity (compression in time) as it will result
in many phase jumps due to the randomisation. The method may also
result in the introduction or modulation of amplitudes over timescales
of the order of, or longer than, the frame length, which while visible
in the waveforms are not noticeably audible typically. Since random
numbers are applied to the phase, this method would only be invertible
if those random numbers were saved as part of the process. Nonetheless,
like the phase vocoder, PaulStretch leaves the STFT spectrogram invariant.

\subsubsection{Wavelet phase vocoder}

\begin{figure}[tbph]
\begin{centering}
\includegraphics{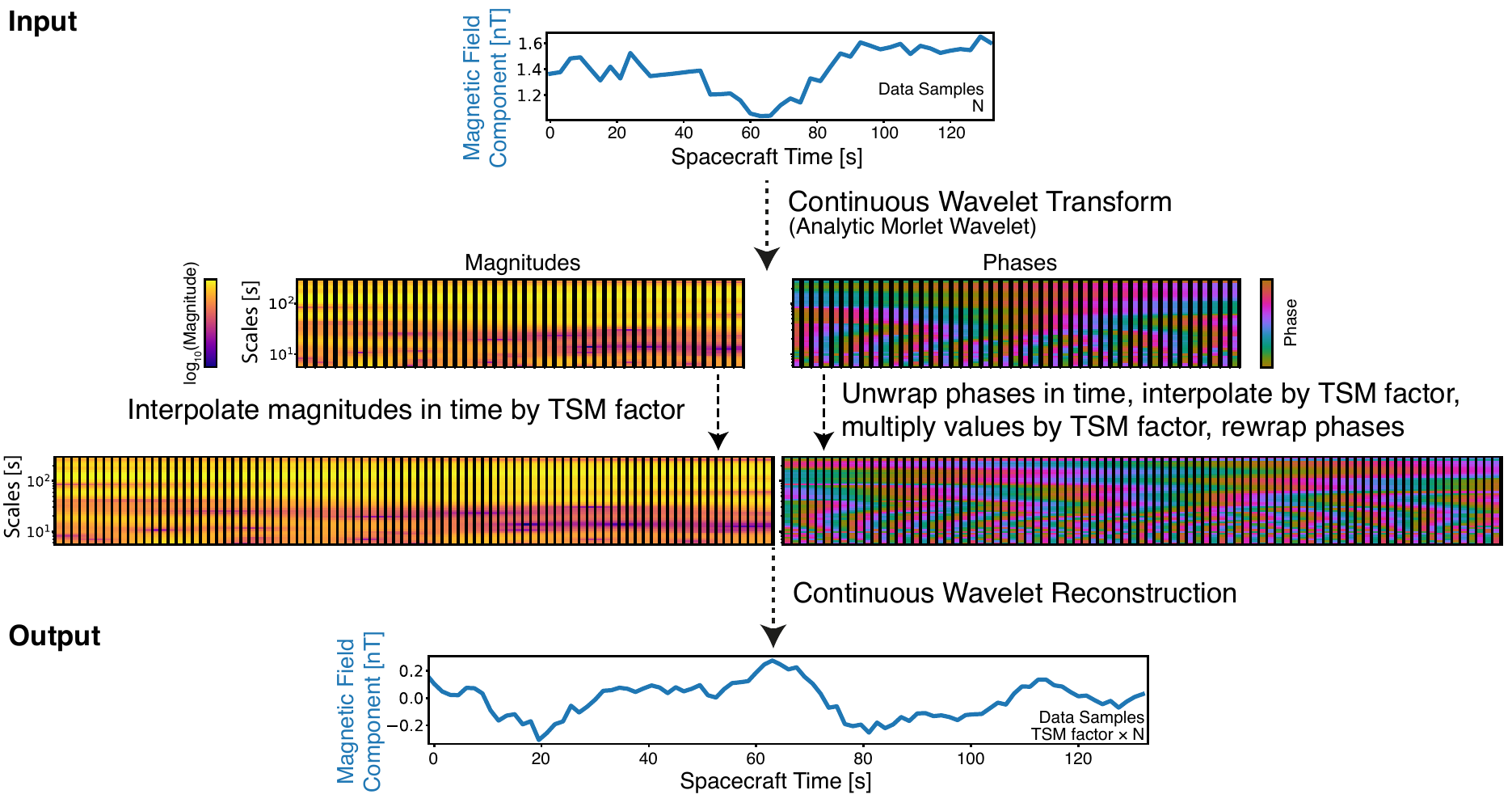}
\par\end{centering}
\caption{Illustration of the Wavelet phase vocoder method. Note the shorter
time range presented due to the different method. \label{fig:wavelet}}

\end{figure}

The wavelet phase vocoder technique (henceforth simply wavelets) works
rather differently to the previous overlap-add procedures, as illustrated
in Figure~\ref{fig:wavelet}. It utilises a complex continuous wavelet
transform of the data \citep{torrence98}, a complete time-frequency
(over) representation of the data which consistently scales these
two dimensions with one another (unlike the STFT). This has the benefit
that magnitude and phase are provided at each frequency for all sampling
times, rather than only at a small number of specified analysis frames.
The TSM procedure is simply an interpolation in time of the wavelet
coefficients (essentially different bandpasses) followed by multiplying
the unwrapped phase by the TSM factor \citep{degersem97}. The interpolation
step increases the number of samples, lowering the pitch of oscillations,
which is subsequently corrected by rescaling the phases by the same
factor. The modified wavelet coefficients are then inverted back to
a time-series \citep{torrence98}. While wavelet reconstruction is
possible mathematically for continuous functions, it has been noted
that this is often not perfect in practice computationally \citep{lebedeva14,postnikov16}.
The example of inversion in Figure~S1 (yellow) shows a constant phase
offset has resulted, likely due to edge effects, but otherwise the
wavelet phase vocoder performs somewat similarly to its STFT counterpart.
The wavelet spectrogram is invariant under this TSM method.

\subsection{Spectral whitening}

The amplitudes of ULF wave magnetic field oscillations in general
are largest at the lowest frequencies and tend to decrease rapidly
with increasing frequency. This pattern is associated mostly with
incoherent background noise, whose overall levels vary depending on
driving and magnetospheric conditions, on top of which discrete resonances
may also be present. However, because of this trend, spectrograms
tend to show undue prominence to the lowest frequencies, making variations
present at higher frequencies hard to discern. Therefore, it is common
to pre-whiten ULF wave spectra so that the background spectrum is
flat. A simple way to achieve this is through taking the time derivative
of the time-series before calculating spectra, since the amplitudes
approximately follow a $1/f$ Brownian/red noise relationship \citep[e.g.][]{engebretson86,russell99}.
Such spectral whitening may also be helpful for similar reasons in
the sonification of ULF waves. Indeed, \citet{archer18} produced
audifications of both the original and time-derivatives of the data.
Here the whitening step is optionally undertaken after TSM, since
it was found that applying it beforehand had adverse effects on the
time stretching.

\subsection{Audification}

The final step is audification. To ensure that the audio waveforms
are constrained to the dimensionless range -1 to +1, the data is normalised
by the maximum absolute value. This normalised data is then written
to audio using a standard audio sampling frequency of 44,100~Hz,
which as discussed earlier renders most of the ULF wave bands audible
since THEMIS measurements are made every $3\,\mathrm{s}$ (Figure~\ref{eq:time-audification}).
While \citet{archer18} produced separate mono tracks for each component
of the magnetic field and a combined stereo file containing all three
components, here we only produce the separate files, focusing in particular
on the azimuthal component of the magnetic field as this is most relevant
to the Alfv\'{e}n continuum \citep{southwood74}. The audio is encoded
in Waveform Audio File Format (WAV), an uncompressed and the most
common audio format. \citet{archer18} had used Ogg Vorbis compression,
which is near-lossless thereby reducing the file size efficiently,
however, we found that not all applications were able to robustly
work with this format (issues with the MP3 format introducing silence
to the audio eliminating the ability to relate audio time to spacecraft
time were highlighted by \citealp{archer18}).

\section{Public dialogue\label{sec:Public-dialogue}}

It was felt that in order to provide recommendations to ULF wave researchers
on the best method of rendering these waves audible that we should
seek input from various stakeholder communities outside of the space
sciences. This is because these communities either have expertise
in audio and its usage, or form intended target audiences for the
sonified ULF waves. We therefore undertook a public dialogue on the
different sonification methods presented. The study gained ethical
approval through Imperial College London's Science Engineering Technology
Research Ethics Committee process (reference 21IC7145).

\subsection{Methods}

\begin{figure}[tbph]
\begin{centering}
\includegraphics{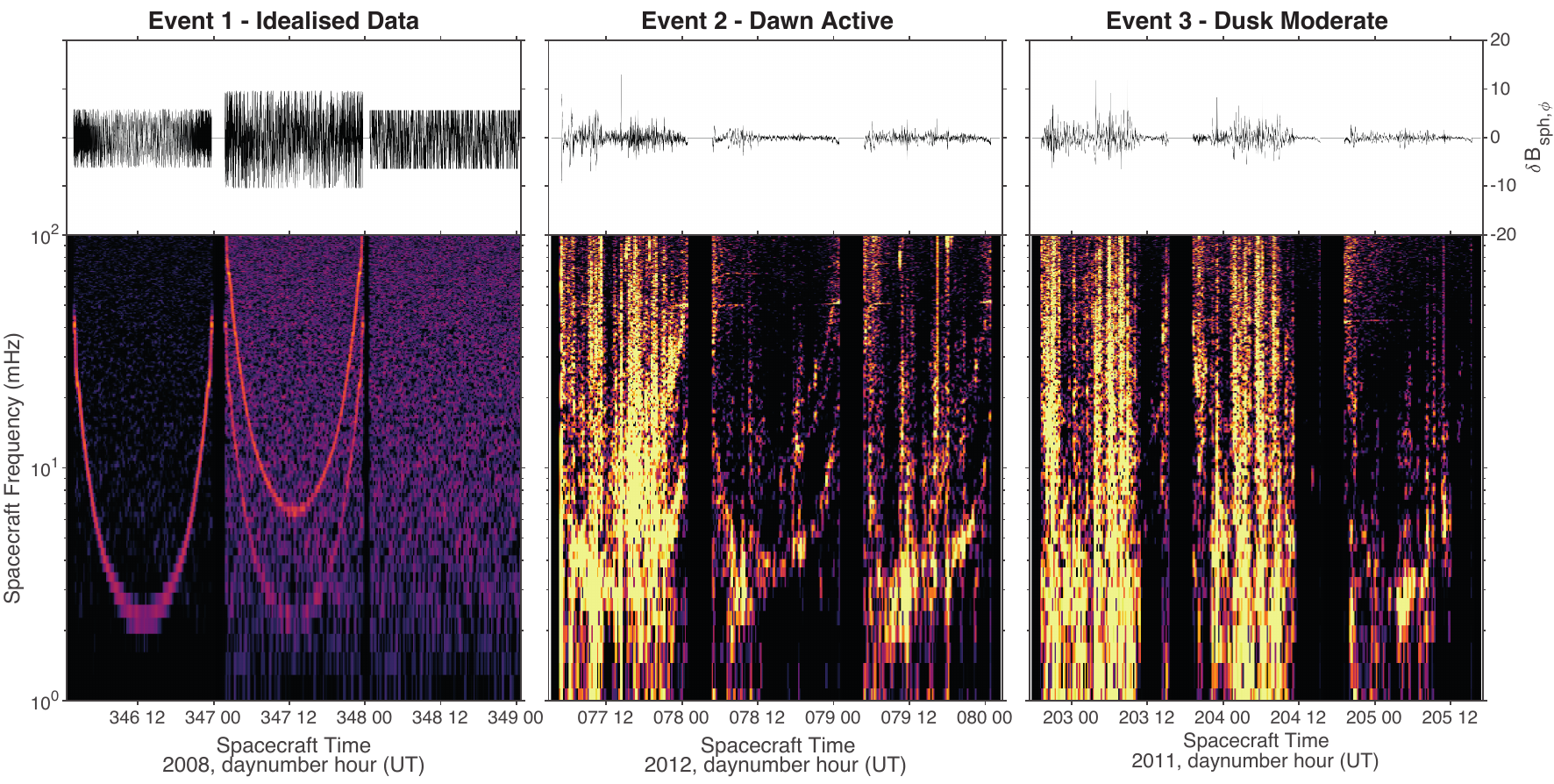}
\par\end{centering}
\caption{The three example THEMIS ULF wave events used in the survey. Top panels
show the azimuthal component of the magnetic field with bottom panels
showing its Short Time Fourier Transform spectrograms with a logarithmic
colour scale, where the background has been spectrally whitened.\label{fig:events}}

\end{figure}

An anonymous online survey was used for the public dialogue, where
participants were asked to rank audio clips and explain their reasoning.
Survey questions can be found in Table~S1. The audio clips in each
question varied either the TSM method, TSM factor, or background noise
spectrum. These were applied to three different THEMIS events. It
was deemed that having more than three events would limit participation,
due to the amount of time it would take to complete the survey, and
that the events chosen provided a good range in ULF waves to apply
the different methods to. Each event corresponds to 3 full orbits
of the THEMIS-E spacecraft starting and ending at perigees, and thus
are approximately 3~days in duration. They are shown in Figure~\ref{fig:events}.
Event~1 corresponds to a synthetic Alfv\'{e}n continuum -- a constant
amplitude chirp/sweep signal, i.e. $\sin\left(\Psi\left[t\right]\right)$
with instantaneous frequency $f\left[t\right]=\partial_{t}\Psi\left[t\right]/2\pi$.
The frequencies are taken from the statistical Alfv\'{e}n continuum
calculations of \citet{archer15b,archer17} based on a large database
of THEMIS density observations. The average frequencies in the dawn
sector as a function of $L$-shell (neglecting plasmaspheric densities
and assuming local azimuthal symmetry) have been constructed and then
mapped to a representative orbit. The first orbit consists of the
fundamental frequency plus low-level (standard deviation of 10\% the
chirp amplitude) Brownian noise \citep{gardiner09}; the second orbit
is a superposition of the first and third harmonics (each at half
the previous amplitude) plus more intense noise (3 times greater);
and the third orbit is just the high-level of noise. The other two
events correspond to real data under different conditions. Event 2
covers 17--20 March 2012, where THEMIS was in the dawn sector (05:40
MLT at apogee) and geomagnetic conditions were active (minimum Dst
was $-44\,\mathrm{nT}$). Event 3 covers 21--24 July 2011, where
THEMIS was in the dusk sector (19:55 MLT at apogee) and geomagnetic
conditions were moderate (minimum Dst was $-20\,\mathrm{nT}$). All
the audio clips used in the survey can be found in Supplementary~Data~2
of \citet{suppdata} or in the survey preview link provided.

\subsubsection{Participants}

Participants were recruited by advertising the study to public email
lists for those with relevant interests/expertise, such as music,
citizen science, or science communication. Social media posts were
also used. A total of 140 people completed the survey over the month
it was open, all of whom indicated their consent as part of the survey
itself. A breakdown of participants' self-identified expertise, shown
in Figure~S2, reveals we successfully targeted the survey outside
of the space science community, with a good mix of interest groups.

\subsubsection{Analysis}

Both quantitative and qualitative analyses are employed, as the closed
and open questions in the survey generate different types of data.

The quantitative data comes in the form of rankings, where a rank
of 1 corresponds to the most liked/favoured (i.e. the highest ranked)
and $r_{max}$ corresponds to the least liked/favoured (i.e. the lowest
ranked). The proportions of all responses $p_{i}$ in each rank $r_{i}$
is determined. Based on these, for each audio clip we construct a
score calculated as
\begin{equation}
\mathrm{Score}=\sum_{i=1}^{r_{max}}\frac{r_{max}-2r_{i}+1}{r_{max}-1}p_{i}
\end{equation}
where the fraction normalises each rank value to between -1 (least
liked) and +1 (most liked), hence the score is also constrained to
this range. The scores are then averaged over the three events to
give an overall score for that option. 95\% confidence intervals in
these overall scores are estimated by bootstrapping the same calculations
over the participants \citep{efron93}, where 5,000 different random
samples with replacement are employed.

Thematic analysis \citep{Braun2006} is used to analyse the meaning
behind the open-text questions. This finds patterns, known as qualitative
codes, in the data which are then grouped into broader related themes.
The themes and codes are defined by induction, where they are iteratively
determined by going through samples of the qualitative data rather
than being pre-defined \citep{Silverman2010,Robson2011}. Once finalised,
the themes and codes are applied to the full dataset by the primary
coder and indicative quotes are identified. A second coder independently
analysed a subset of the data (30 participants' responses, corresponding
to 21\%) to check reliability. Average agreement across the different
themes ranged from 73--95\%. A typical measure of reliability in
coding qualitative data is Cohen's kappa, a statistic between 0 and
1, given by unity minus the ratio of observed disagreement to that
expected by chance \citep{mchugh12}. Average values between 0.5--0.6
were found across the themes, which correspond to 70--80\% of the
maximum achievable values given the uneven distributions of the data
\citep{umesh89}. All these statistics therefore indicate good agreement
and thus the qualitative analysis is reliable.

\subsection{Results}

\begin{figure}[tbph]
\begin{centering}
\includegraphics{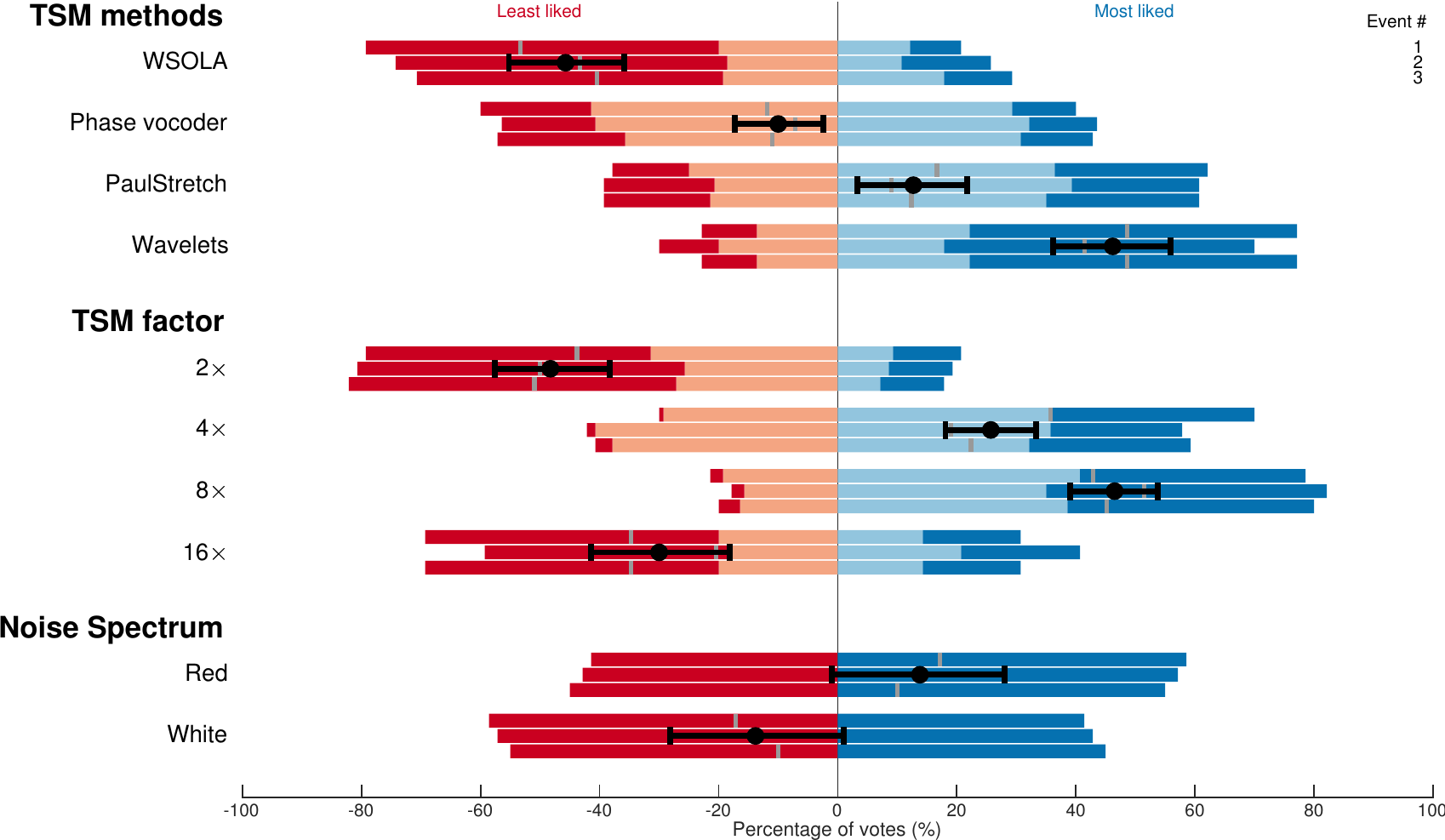}
\par\end{centering}
\caption{Quantitative results of the survey. Stacked bars show the proportions
of each ranking from least liked (dark red, far left) to most liked
(dark blue, far right). The score for each audio clip is indicated
by the grey bars, with the overall score across all 3 clips for each
group shown as the black marker along with its 95\% confidence interval.
Note that for TSM methods and TSM factors there were four possible
options, and thus also four possible ranks, while for noise spectrum
there were only two.\label{fig:survey-results}}

\end{figure}

All participants' responses to the survey question can be found in
Supplementary~Data~3. The quantitative results of the survey are
shown in Figure~\ref{fig:survey-results}, depicting the proportions
in each rank for the various options as well as their scores. No obvious
trends were present in the quantitative results between different
interest groups, hence we simply present all the data together in
this paper. Table~\ref{tab:themes} summarises the themes and their
underlying codes that were determined from the qualitative data, as
well as which question topics these pertained to. The application
of these qualitative codes to the open responses can be seen in Supplementary~Data~4
of \citet{suppdata}. One of the main themes was \textit{timbre},
which refers to the perceived quality of a sound. This theme encapsulates
aspects of whether the sounds were pleasant to listen to or if potential
issues around ear fatigue were raised. The codes in this theme thus
are either positive, negative, or neutral. The second theme concerned
\textit{signal-to-noise} or the perceived information content within
the sounds, i.e. whether the tones were discernible from the background.
Again the codes ranged from positive, through to neutral, or negative.
Issues around potential \textit{artifacts} introduced by the processing
were raised as another theme. Many respondents also commented with
\textit{synonymous sounds}, i.e. what they thought the audio ``sounded
like''. The final themes/codes concerned whether it was possible
to hear the \textit{detail} present, if listening to the sounds evoked
a sense of \textit{boredom}, and if the participants desired more
\textit{context} on the intended usage of the sounds. All these themes
were present across the different interest groups. Not every participant
answered all the open-text questions, with response rates between
$88\text{--}94\%$. Some answers did not fit into all of the themes
either, with the two main themes of timbre and signal-to-noise being
discussed on average in 41\% and 30\% of responses respectively.

\subsubsection{TSM method}

Figure~\ref{fig:survey-results} indicates that the wavelets technique
was by far the participants' favourite TSM method. Its timbre was
deemed to be the most pleasant to the ear, with 57 positive responses
compared to only 5 neutral and 3 negative
\begin{quote}
``The best on all counts. Good depth, and richness in texture.''
(Participant~3)\\
``This had a softer sound, easier to listen to.'' (Participant~41)
\end{quote}
Many participants (a total of 37) noted that the results of this method
evoked the sounds of water. In terms of the signal-to-noise, 26 responses
indicated that the tones were sufficiently clear
\begin{quote}
``Conveyed the frequencies effectively.'' (Participant~6)\\
``Clearly isolates the components of the signals.'' (Participant~56)
\end{quote}
Only 7 expressed the wavelets method was too noisy, though 16 responses
indicated more neutral responses within this theme
\begin{quote}
``Very `harmonic' sounds, but occasionally hard to differentiate.''
(Participant~45)
\end{quote}

PaulStretch had the second highest overall score. Generally it's timbre
was thought of as positive (32 responses)
\begin{quote}
``I like it, makes sounds very smooth and kind of diffuse.'' (Participant~15)\\
``This was also pleasing, but a little less than Wavelets.'' (Participant~84)
\end{quote}
but more neutral (14) and negative (13) comments were made than with
wavelets
\begin{quote}
``Kind of a middle ground between the watery feel of Wavelets and
the glitchy techno of the other two'' (Participant~14)\\
``Felt very static filled and hard to listen to.'' (Participant~128)
\end{quote}
The most common synonymous sounds were those of wind or ``natural''
sounds. The survey results were inconclusive on the signal-to-noise
ratio present with PaulStretch.

The phase vocoder method was ranked only slightly below PaulStretch
-- several participants noted similarities in the sounds of the two,
which is due to their related methods. However, open responses were
more negative (36 responses) on how this method sounded
\begin{quote}
``Sounds like really terrible radio interference and is very jarring
to the nerves.'' (Participant~35)\\
``The metallic character made it less pleasant to listen to'' (Participant~25)
\end{quote}
compared to 17 neutral and 6 positive comments. Results were again
mixed on the information content, however, potential artifacts associated
with this method were more commonly raised than before (12 responses)
\begin{quote}
``This sounds like heavily processed noise cancelling DSP {[}digital
signal processessing{]} which maybe good at recovering spoken words
but heavily masks fundamental random signals.'' (Participant~55)\\
``Most unpleasant; phasing is the culprit.'' (Participant~88)
\end{quote}

WSOLA was clearly the least liked TSM method in the rankings. Indeed,
almost all comments on the sound quality were negative (70 responses)
\begin{quote}
``Too much distortion for my taste, hard for me to listen to.''
(Participant~86)\\
``Totally unlistenable. It sounds like 4-bit digital audio.'' (Participant~129)\\
``Lack of amplitude dynamic{[}s{]} makes it almost painful to listen.''
(Participant~135)
\end{quote}
Similar to with the phase vocoder, results on the signal-to-noise
were inconclusive and processing artifacts were raised several times
(17 responses)
\begin{quote}
``Distortion artifacts are probably not real'' (Participant~56)\\
``Too much artificial noise sounds.'' (Participant~103)
\end{quote}

Therefore, the recommendation from our survey is that the wavelet
phase vocoder method is the preferred TSM method for application to
magnetospheric ULF waves. It may be possible to improve this method
even further by compensating for the spreading effects in time-frequency
due to the mother wavelet. Examples of this are the synchrosqueezed
wavelet transform \citep{daubechies11} which uses reassignment in
frequency, or superlets \citep{moca21} which is a geometric combination
of sets of wavelets with different bandwidths. Further work into how
one may apply these to TSM is required.

\subsubsection{TSM factor}

The result of the rankings shown in Figure~\ref{fig:survey-results}
shows that a TSM factor of $8\times$ was favoured, with $4\times$
somewhat close behind (this difference is statistically significant
though). TSM factors of $2\times$ and $16\times$ were both ranked
poorly and the confidence intervals in their overall scores overlap.
The reason behind these scores could be gleaned from the open-text
responses. Participants stated that larger TSM factors allowed more
time to hear the detail of the signals within the clips, which was
not possible with the shortest clips (44 responses)
\begin{quote}
``Length of audio clips coincided with perception of individual tones
and increased clarity of sounds as the length increased.'' (Participant~5)\\
``I much prefer the longer audio lengths, because it allows me to
hear the nuances in the received sound.'' (Participant~78)\\
``The 2x signal goes by too fast. Listener will miss small changes
in the signal.'' (Participant~59)
\end{quote}
However, in contrast, it was felt that the longest clips may induce
boredom in the listener (30 responses)
\begin{quote}
``While having a 30 second clip would be ideal, it feels tedious
and boring and not `fun' to listen to. It also can be quite painful
to listen to some of the tracks at full length.'' (Participant~5)\\
``Generally the $16\times$ feels dragging too slow and information
can be obtained from faster speeds.'' (Participant~20)
\end{quote}
Thus the consensus was that the two middle options provided a compromise
to both these themes, though individuals' preferences varied between
$4\times$ and $8\times$. 11 participants raised that to best answer
the question on the TSM factor they would have preferred to know more
about the context of the sounds, their intended uses, and any tasks
associated with them. We intentionally did not provide this, however,
as our aim was to arrive at broad recommendations on the sonification
of magnetospheric ULF waves that may be applied in a variety of contexts
and settings.

The recommendation on TSM factor from the survey would be to use a
value of $\sim6\times$, based on the average (either arithmetic or
geometric) TSM factors for $4\times$ and $8\times$ using the overall
scores as weights.

\subsubsection{Noise spectrum}

Participants' preferences in terms of the background noise spectrum
of the audio were somewhat split, as shown in Figure~\ref{fig:survey-results},
with overall 57\% preferring red noise (audification of $\mathbf{B}$)
to white noise ($\partial_{t}\mathbf{B}$). The confidence intervals
for the scores also are not constrained to simply positive or negative.
Therefore, the quantitative results do not provide a clear recommendation.
However, the qualitative data provides further insight. While opinions
on which had the more pleasant timbre were again somewhat split, comments
on the harshness of the white noise (23 responses) outweighed those
of the red (8 responses) with many references to the higher frequencies
present being the cause of this
\begin{quote}
``The white noise has higher frequencies, which is giving me some
ear fatigue.'' (Participant~19)\\
``This noise hurts my ears and gives me a headache. It is sharper
and tinnier.'' (Participant~26)
\end{quote}
However, it was recognised that the spectral whitening made it easier
to distinguish the signals (30 responses)
\begin{quote}
``The tones seemed clearer against this as a backdrop.'' (Participant~21)\\
``This process provides a full spectrum appreciation of the underlying
signals that are not bandwidth limited due to masking or filtering.''
(Participant~55)
\end{quote}
whereas the red noise sounded somewhat ``muffled'' and less clear.

The spectral whitening of ULF waves is used to make signals over a
wide range of frequencies clearer in spectrograms, since the background
spectrum becomes approximately constant with frequency \citep{engebretson86,russell99}.
The same power value in the spectrogram relative to the background
at different frequencies therefore can be seen as the same colour
on the chosen colourmap. However, the survey results highlight that
the same level of intensity of sound at different frequencies are
not perceived as the same loudness. Indeed, human hearing is most
sensitive to higher frequencies in the range $2\text{--}5\,\mathrm{kHz}$,
which is likely the reason for the comments on the harshness of the
white noise. \textit{Equal-loudness contours} have been determined,
which specify what sound pressure levels at different frequencies
are perceived as being at the same loudness level \citep{ISO226,suzuki04}.
Therefore, rather than modifying the spectrum of the ULF waves for
sonification to be flat in intensity they should be adjusted for equal-loudness,
which should be possible through applying appropriate filtering (e.g.
by modifying the magnitudes of the Fourier transform after stretching).
This should then have the benefits of making tones discernible but
not being too harsh on the ears.

\section{Discussion}

While time-series data of magnetospheric ultra-low frequency (ULF)
waves are often still analysed visually (at least in part), this form
of data lends itself more naturally to our sense of sound. Direct
audification is the simplest sonification method, providing a true
representation of the original data. When applied to ULF waves though
this can result in changes occurring too rapidly for effective analysis
by the human auditory system. Therefore, we detail several existing
audio time scale modification (TSM) techniques which have been applied
to ULF wave data. Through a public dialogue with stakeholder groups,
we arrive at recommendations on which sonification methods should
be used to best render the Alfv\'{e}n waves present audible, which
are summarised in Table~\ref{tab:recommendations}. We have implemented
these final recommendations, applying them to the three THEMIS example
events yielding the audio in Supplementary~Data~5 of \citet{suppdata}.

Figure~\ref{fig:events} shows a typical spectrogram representation
of ULF waves for the three examples, where the logarithmic colour
scale has been spectrally whitened and the limits of the colour scale
in each individual event have been set at the $50\%$ (corresponding
to the noise level) and $95\%$ (corresponding to the peaks) percentiles
in power in order to capture the range present. In the idealised data
(event~1) the Alfv\'{e}n continuum, with frequencies decreasing
from perigee to apogee, is very clear. In contrast, in the real data
(events~2--3) identifying discrete peaks even by visual inspection
is much more difficult. Even in the dawn sector under active geomagnetic
conditions, where standing toroidal Alfv\'{e}n waves are more common
and their frequency profiles should be simpler (i.e. similar to event~1;
\citealp{takahashi16,archer17}), the continuum is still subtle, especially
for the first orbit where significant incoherent broadband wave power
is also clearly present. In contrast, the Alfv\'{e}n continuum is
clearly audible in Supplementary~Data~5 \citep{suppdata} throughout
portions of the orbits for all three events. The auditory system's
ability to identify these subtle sweeping frequency tones in the presence
of significant noise and other potential signals is likely thanks
to its nonlinear nature and impressive ability at blind source separation.
It is well known that all wave analysis techniques have their advantages
and drawbacks, which will depend on the nature of the precise oscillations
present \citep{chi08,piersanti18}. Sonification can thus provide
an additional supportive tool for researchers in identifying different
ULF waves \citep{alexander11,alexander14,wicks16,archer18} that may
complement other techniques. By maximising the audibility of ULF waves
for more challenging orbits/environments/events, the methods presented
here should hopefully improve further the utility of sonification
in this science topic.

Another benefit to sonification is that it renders scientific data
more accessible and lowers the barrier to entry for students and the
public to contribute to space science through citizen science \citep{archer18}.
With the growing number of space plasma spacecraft in orbit around
Earth and the networks of ground magnetometers globally, we are continually
producing \textit{big data} that poses a challenge to efficiently
navigate, mine, and analyse. Machine learning is typically the emerging
solution to dealing with big data in general, with supervised machine
learning techniques being applied to a variety of space physics tasks
\citep[e.g.][]{10.3389/fspas.2020.00055,https://doi.org/10.1029/2020EA001530}.
However, current challenges in ULF wave research mean that many simple
tasks (e.g. classifying ULF wave events) are still not easily tackled
by these methods due to the lack of good (e.g. classified) training
sets of events. Until ULF wave research can be fully automated, clearly
it is not feasible for a single researcher to visually inspect all
the ULF wave data that is being produced. The dramatically reduced
analysis processing time associated with listening to sonified data
(even with moderate TSM applied) certainly helps. However, any manual
process applied by a single researcher is potentially subject to biases
and concerns over reliability. On the other hand, mobilising citizen
scientists en masse to cover these vast datasets and arrive at a statistical
consensus for each interval/event may in fact be more robust \citep[e.g.][]{barnard14}.
Therefore, there is a lot of potential in applying the sonification
methods presented here to arrive at new scientific results through
citizen science. A simple example is that already discussed, identifying
how the properties and excitation of the Alfv\'{e}n continuum vary
under different solar wind and geomagnetic driving conditions --
an important and still unresolved issue \citep[e.g.][]{rae19}. Another
possibility is that citizen scientists collectively may be able to
arrive at more data-driven classifications of ULF waves that take
into account further properties of the waves than simply frequency,
which may better distinguish between the different physical processes
at play than the current scheme \citep{jacobs64}. Countless other
scientific questions into the sources and propagation of ULF waves
in planetary magnetospheres could be addressed through citizen science
with sonified data. Indeed the pilot ``Heliophysics Audified: Resonances
in Plasmas'' (HARP) citizen science project (\url{http://listen.spacescience.org/})
is already building on the work presented by \citet{archer18} in
this area, developing more streamlined interfaces for citizen scientists
to interact with the audible data and record scientific results. A
result of increased citizen science in ULF wave research could be
the very training sets required to be able to apply machine learning
algorithms to the data, an approach which has successfully been done
in other fields \citep[e.g.][]{Beaumont_2014,SULLIVAN201431}. Once
trained, these machine learning algorithms would then be able to tie
together multi-satellite and multi-station data of the same event
at different locations, in ways which are not possible with a single
audio stream, to improve our global understanding of system-scale
magnetospheric dynamics under different driving regimes.

More work is required to understand the full scope of sonification
in the identification, categorisation, and characterisation of the
zoo of ULF waves present within Earth's magnetosphere. The application
of existing TSM methods from the field of music and audio in this
paper was motivated by the short timescales associated with direct
audification of ULF waves and limits in human's pitch perception based
on the number of oscillations in typical ULF wave events. While this
work has certainly increased the audibility of ULF waves, the Alfv\'{e}n
continuum in particular, only through further work in applying sonification
for the purposes of novel scientific results can the full benefits
and limits of these tools be realised.

Beyond potential scientific benefits, there are also obvious uses
of sonified ULF waves in education, engagement, and communication.
Recently a number of high-profile ULF wave results have leveraged
the methods presented in this paper within press releases for the
media \citep{ncei18,nasadrum19,esaforeshock19,nasastanding21}, which
have gone on to successfully attract global attention. Therefore,
sonification is a helpful tool in communicating our science. \citet{archer_soundscape21}
showed that simply enabling public audiences to experience these sounds
can spark innate associations and dispell common misconceptions simply
through the act of listening, highlighting the power of the medium
in its own right. This has similarly been reflected in many of the
survey responses of synonymous sounds, e.g. the perceived water-like
quality of the wavelets processed data may spark conversations about
fluids and (magneto)hydrodynamics in space. More in-depth engagement
projects that enable high school students to work with audible data
as part of research projects \citep{archer18,archer_prise_framework21}
have recently been shown to have immense benefits to students, teachers
and schools from a variety of backgrounds \citep{archer_prise_impact21,archer_prise_schools21}.
These include increased confidence, developed skills, raised aspirations,
and greater uptake of science. Sonifications may also be used as creative
elements in the production of art, thereby engaging those who might
not actively seek out science otherwise \citep{archer_ssfx,poem21}.
Therefore, the potential uses of these methods are vast.

Finally, the sonification methods beyond direct audification presented
here could easily be applied to other forms of waves. Indeed, there
is a long history of converting heliophysics data across different
frequency bands into audible sounds. The terminology of ionospheric
extremely-low frequency (ELF) and very-low frequency (VLF) radio waves,
which already span the human hearing range, were largely based on
on their psychoacoustics when picked up by radio antenna, e.g. ``whistlers''
\citep{barkhausen19} and ``lion roars'' \citep{smith67}. This
tradition has continued with terms such as ``tweaks'', ``chorus'',
``hiss'' and ``static'' being commonly used across heliospheric
research. Many examples of such higher (than ULF) frequency waves
from across the solar system, either already in the audible range
or in fact pitched down to be rendered audible, are available online
(e.g. \href{http://space-audio.org/}{http://space-audio.org/}). While
the specific recommendations (such as the TSM factor and audio sampling
rate) made here are tailored for the Pc3--6 ULF wave bands, and the
Alfv\'{e}n continuum in particular, there is no reason why these
choices could not be suitably adjusted for other waves/frequencies
to improve their audibility also. For example, electromagnetic ion
cyclotron (EMIC) waves typically are found in the Pc1--2 ULF wave
bands and would require different choices of parameters to render
them audible. Even electron cyclotron harmonic (ECH) waves, which
already occupy the audible range, can benefit from some TSM (see an
example in \citealp{spaceweatherech21}). There are clearly also applications
to time-series data in general, not just within the space sciences.
Therefore, there are potentially many ways that the scientific community
and wider society can benefit from this work into sonification.

\section*{Conflict of Interest Statement}

The authors declare that the research was conducted in the absence
of any commercial or financial relationships that could be construed
as a potential conflict of interest.

\section*{Author Contributions}

MOA and MDH conceived of the study and supervised the project. MC,
XS, SC, and EDH contributed to the sonification method design and
software development. MOA, MDH, XS, MFJF, and EVM contributed to the
survey design. MOA and MFJF performed the analysis. MOA wrote the
manuscript. All authors contributed to manuscript revision, read,
and approved the submitted version.

\section*{Funding}

MOA holds a UKRI (STFC / EPSRC) Stephen Hawking Fellowship EP/T01735X/1.
MC was funded by The Ogden Trust physics education grant PEGSU21\textbackslash 101
through Imperial College London's Undergraduate Research Opportunity
Programme. MDH was supported by NASA grant 80NSSC21K0796 and NASA
grant 80NSSC19K0907. XS is supported by NASA award 80NSSC19K0907.

\section*{Acknowledgments}

MOA would like to thank Sophia Laouici, Takudzwa Makoni, and Nivraj
Chana, whose preliminary work through undergraduate summer internships
funded by The Ogden Trust ultimately helped guide this study. We acknowledge
NASA contract NAS5-02099 and V. Angelopoulos for use of data from
the THEMIS Mission. Specifically: C. W. Carlson and J. P. McFadden
for use of ESA data; K. H. Glassmeier, U. Auster and W. Baumjohann
for the use of FGM data provided under the lead of the Technical University
of Braunschweig and with financial support through the German Ministry
for Economy and Technology and the German Center for Aviation and
Space (DLR) under contract 50 OC 0302.

\section*{Supplementary Material}

\noindent\textbf{Supplementary~Table~1} The survey questions.

\noindent\textbf{Supplementary~Figure~1} Examples demonstrating
the degree of reversibility of TSM methods in practice.

\noindent\textbf{Supplementary~Figure~2} Venn diagram of survey
participants' self-identified expertise.

\section*{Data Availability Statement}

The THEMIS data was obtained and processed using the software developed
in this study, which is available at \url{https://github.com/Shirling-VT/HARP_sonification}.
The survey and its embedded audio clips can be previewed at \url{https://imperial.eu.qualtrics.com/jfe/preview/SV_295iuL4yxfaQ0Qu?Q_CHL=preview&Q_SurveyVersionID=current}.
All other data used is available in \citet{suppdata}.

\bibliographystyle{frontiersinSCNS_ENG_HUMS}
\bibliography{timestretch}

\clearpage{}

\section*{Tables}

\begin{table}[H]
\centering{}%
\begin{tabular}{|c|c|c|}
\hline 
\textbf{Question subject(s)} & \textbf{Theme} & \textbf{Codes}\tabularnewline
\hline 
\hline 
\multirow{6}{*}{TSM method \& noise spectrum} & \multirow{3}{*}{Timbre} & Positive\tabularnewline
\cline{3-3} 
 &  & Neutral\tabularnewline
\cline{3-3} 
 &  & Negative\tabularnewline
\cline{2-3} \cline{3-3} 
 & \multirow{3}{*}{Signal-to-noise} & Positive\tabularnewline
\cline{3-3} 
 &  & Neutral\tabularnewline
\cline{3-3} 
 &  & Negative\tabularnewline
\hline 
\multirow{2}{*}{TSM method} & \multicolumn{2}{c|}{Artifacts}\tabularnewline
\cline{2-3} \cline{3-3} 
 & \multicolumn{2}{c|}{Synonymous sounds}\tabularnewline
\hline 
\multirow{3}{*}{TSM factor} & \multicolumn{2}{c|}{Detail}\tabularnewline
\cline{2-3} \cline{3-3} 
 & \multicolumn{2}{c|}{Boredom}\tabularnewline
\cline{2-3} \cline{3-3} 
 & \multicolumn{2}{c|}{Context}\tabularnewline
\hline 
\end{tabular}\caption{Themes and codes from the qualitative data.\label{tab:themes}}
\end{table}

\begin{table}[H]
\centering{}%
\begin{tabular}{|c|c|}
\hline 
\textbf{Sonification choice} & \textbf{Recommendation}\tabularnewline
\hline 
TSM method & Wavelet phase vocoder\tabularnewline
TSM factor & $6\times$\tabularnewline
Noise spectrum & Equal-loudness contour\tabularnewline
Audio sampling rate & 44,100~Hz\tabularnewline
Waveform amplitude & Normalisation per interval\tabularnewline
\hline 
\end{tabular}\caption{Final recommendations on ULF wave sonification.\label{tab:recommendations}}
\end{table}

\section*{\clearpage Figure captions}

\pagebreak{}
\pagebreak{}
\pagebreak{}
\pagebreak{}
\pagebreak{}
\pagebreak{}
\pagebreak{}

\end{document}


\title{Suppementary Material}
\maketitle

\section{Supplementary Data}

\textbf{Supplementary~Data~1 }ZIP folder containing audio files
of the three THEMIS events using direct audification.

\noindent\textbf{Supplementary~Data~2} ZIP folder containing all
audio files embedded within the survey.

\noindent\textbf{Supplementary~Data~3} All participants' responses
to the survey questions.

\noindent\textbf{Supplementary~Data~4} Qualitative coding of the
open responses.

\noindent\textbf{Supplementary~Data~5} ZIP folder containing audio
files of the three THEMIS events using the recommended sonification
methods.

\section{Supplementary Tables and Figures}

\subsection{Tables}

\begin{table}[tbph]
\centering{}\begin{footnotesize}%
\begin{tabular}{|l|>{\raggedright}p{8cm}|>{\raggedright}p{2.5cm}|>{\raggedright}p{4cm}|}
\hline 
\textbf{No.} & \textbf{Question} & \textbf{Response type} & \textbf{Options}\tabularnewline
\hline 
\multirow{5}{*}{1} & \multirow{5}{8cm}{What area(s) of expertise relevant to this project would you say you
have? (tick all that apply)} & \multirow{5}{2.5cm}{Multiple choice (allow multiple)} & Audio / Music / Radio\tabularnewline
\cline{4-4} 
 &  &  & Citizen Science / Crowdsourcing\tabularnewline
\cline{4-4} 
 &  &  & Public Engagement / Science Communication\tabularnewline
\cline{4-4} 
 &  &  & Space Science\tabularnewline
\cline{4-4} 
 &  &  & Other (please specify)\tabularnewline
\hline 
\multirow{4}{*}{2a--c} & \multirow{4}{8cm}{Please drag and drop the following audio clips ordering them from
1--4, where 1 is your favourite and 4 is your least favourite.} & \multirow{4}{2.5cm}{Rank order} & Paulstretch\tabularnewline
\cline{4-4} 
 &  &  & Phase vocoder\tabularnewline
\cline{4-4} 
 &  &  & Wavelets\tabularnewline
\cline{4-4} 
 &  &  & WSOLA\tabularnewline
\hline 
\multirow{4}{*}{2d} & \multirow{4}{8cm}{Briefly describe what you thought of each method. We value your opinion
so there are no right or wrong answers.} & \multirow{4}{2.5cm}{Open text} & Paulstretch\tabularnewline
\cline{4-4} 
 &  &  & Phase vocoder\tabularnewline
\cline{4-4} 
 &  &  & Wavelets\tabularnewline
\cline{4-4} 
 &  &  & WSOLA\tabularnewline
\hline 
\multirow{2}{*}{3a--c} & \multirow{2}{8cm}{Which of the following clips do you prefer?} & \multirow{2}{2.5cm}{Multiple choice (allow one)} & Red noise\tabularnewline
\cline{4-4} 
 &  &  & White noise\tabularnewline
\hline 
\multirow{2}{*}{3d} & \multirow{2}{8cm}{Briefly explain what you thought of each method, again there are no
right or wrong answers.} & \multirow{2}{2.5cm}{Open text} & Red noise\tabularnewline
\cline{4-4} 
 &  &  & White noise\tabularnewline
\hline 
\multirow{4}{*}{4a--c} & \multirow{4}{8cm}{Please drag and drop the following audio clips ordering them from
1--4, where 1 is your favourite and 4 is your least favourite.} & \multirow{4}{2.5cm}{Rank order} & $2\times$\tabularnewline
\cline{4-4} 
 &  &  & $4\times$\tabularnewline
\cline{4-4} 
 &  &  & $8\times$\tabularnewline
\cline{4-4} 
 &  &  & $16\times$\tabularnewline
\hline 
4d & Briefly explain your preference in audio length. There are no right
or wrong answers. & Open text & \tabularnewline
\hline 
\end{tabular}\end{footnotesize}\caption{The survey questions. The survey itself can be previewed at \protect\url{https://imperial.eu.qualtrics.com/jfe/preview/SV_295iuL4yxfaQ0Qu?Q_CHL=preview&Q_SurveyVersionID=current}.}
\end{table}

\subsection{Figures}

\begin{figure}[tbph]
\begin{centering}
\includegraphics{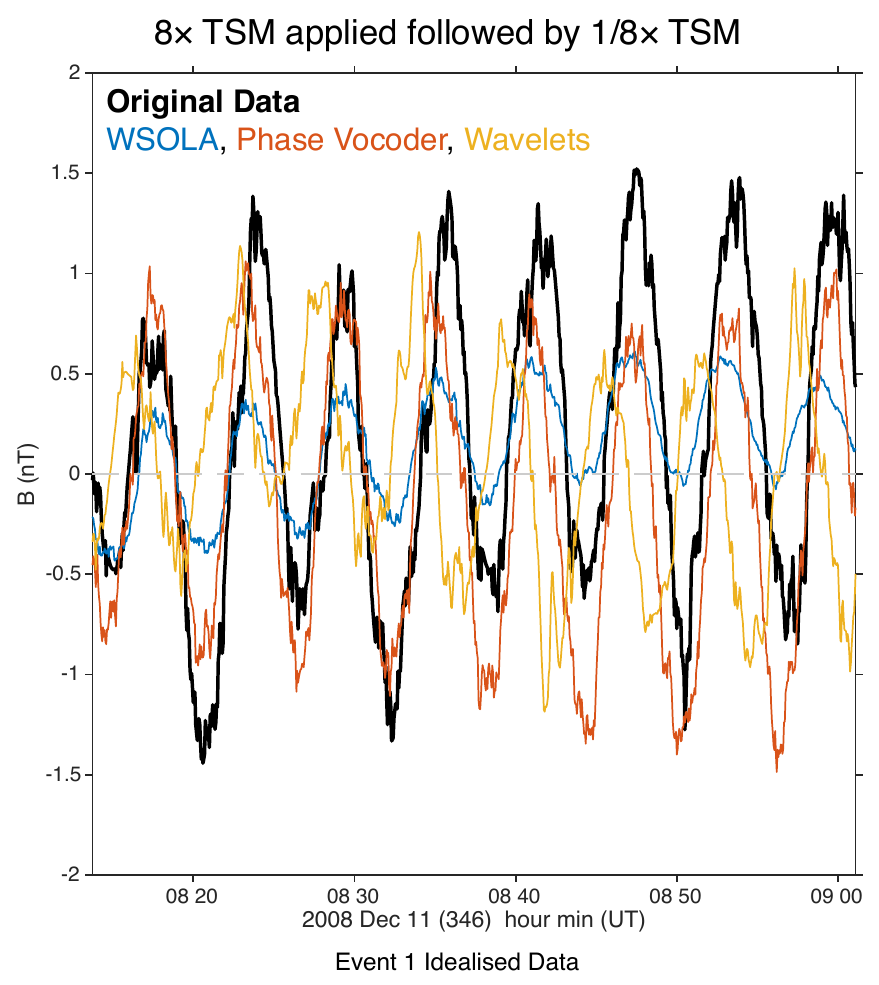}
\par\end{centering}
\caption{Reversability of three of the TSM methods in practice. The original
idealised data (black) is first stretched and then compressed both
through the same methods each time.}

\end{figure}

\begin{figure}[tbph]
\begin{centering}
\includegraphics[width=0.75\columnwidth]{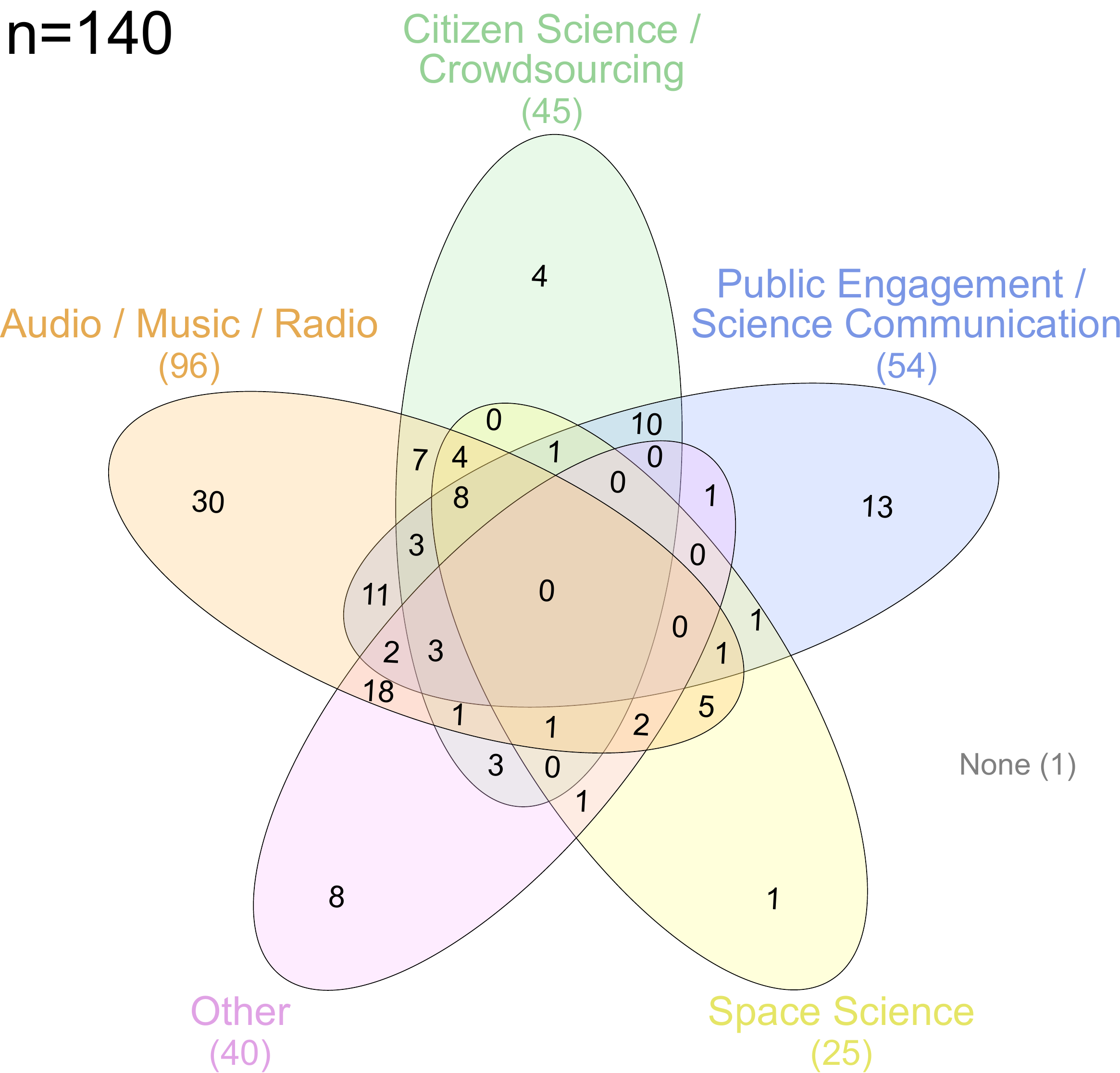}
\par\end{centering}
\caption{Venn diagram of survey participants' self-identified expertise.}

\end{figure}